\documentclass[12pt]{iopart}
\usepackage{epsfig}
\usepackage{amssymb}
\usepackage[dvips]{color}

\newcommand{\be}{\begin{equation}}
\newcommand{\ee}{\end{equation}}
\newcommand{\bea}{\begin{eqnarray}}
\newcommand{\eea}{\end{eqnarray}}
\newcommand{\ba}{\begin{array}}
\newcommand{\ea}{\end{array}}

\begin{document}


\title[Translocation of a polymer chain driven by a dichotomous noise]{Translocation of a polymer chain driven by a dichotomous noise}

\author{Alessandro Fiasconaro$^{1,}$$^{2}$, Juan Jos\'e Mazo$^{1,}$$^2$, Fernando Falo$^{1,}$$^3$}

\address{$^1$ Departamento de F\'isica de la Materia Condensada, Universidad de Zaragoza. 50009 Zaragoza, Spain}
\address{$^2$ Instituto de Ciencia de Materiales de Arag\'on, CSIC-Universidad de Zaragoza. 50009 Zaragoza, Spain}
\address{$^3$ Instituto de Biocomputaci\'on y F\'isica de Sistemas Complejos, Universidad de Zaragoza. 50018 Zaragoza, Spain}

\ead{afiascon@unizar.es}

\date{\today}

\begin{abstract}
We consider the translocation of a one-dimensional polymer through a
pore channel helped by a motor driven by a dichotomous noise with time
exponential correlation. We are interested in the study of the
translocation time, mean velocity and stall force of the system as a
function of the mean driving frequency. We find a monotonous
translocation time, in contrast with the mean velocity which shows a
pronounced maximum at a given frequency. Interestingly, the stall
force shows a nonmonotonic behavior with the presence of a
minimum. The influence of the spring elastic constant to the mean
translocation times and velocities is also presented.
 \end{abstract}

\maketitle

\section{Introduction}
Translocation features of polymers through natural and artificial
pores is a current active research topic in biophysics and
nanotechnology~\cite{KasPNAS96,RMP,NL}. Motivated by many broad
interest experimental results, different models have been introduced
to describe and study in a simple way this and related problems. For
instance, single barrier potentials \cite{Pizz}, as well as flashing
ratchet models \cite{linke}, have been studied to describe the polymer
translocation and polymer transport dynamics. The passage of small
molecules through passive cell channels can be also modeled by
stochastic and rachetlike forces \cite{shulten}. In some cases the
transport phenomena involves not translocation through pores, but also
molecular motors, whose complex action has been recently addressed at
high attention \cite{Bust,Bust09}. In addition, nanotechnological
applications try to emulate the complex biological process related to
the translocation dynamics \cite{mickler, starikov}.

Recently, we have studied different models for the 1d translocation of
a spring-bead polymer helped by a motor using a sinusoidal
force~\cite{fjf-sine}. The introduction of a time dependent driving
force imposes a new time scale on the system, and provides new and
richer phenomenology: for sinusoidal driving, the translocation time
shows an oscillatory behavior as a function of the frequency.

In order to introduce stochasticity in the motor action and motivated
by the relevant role played by dichotomous noise in biological
problems, in this manuscript we consider the case of a polymer driven
by a two-state force: constant force which pushes the polymer chain in
one direction during the activity of the motor, and zero force which
leaves the polymer to diffuse freely otherwise. This pure dichotomous
mechanism constitutes a first approach in describing a machine working
dichotomously between two on-off states~\cite{gomez,fjf-damn}.

The motor modeled in \cite{gomez,fjf-damn} acts during a fixed time,
while the waiting times are exponentially distributed with a mean time
depending on the ATP concentration. In the present work a simpler
dichotomous mechanism which can well point out, by contrast, the
specific behavior of the ATP based machines is studied.

On the other hand, pure dichotomous driving makes sense in the
nanotechnological context as well as in the biophysical one.  In the
first case the passage of a polymer can be induced through a graphene
pore or solid state channeling \cite{han99,luan2010} by applying a
dichotomous force between the two sides of the layer.  In the second
case, the model can describe the translocation of a linear molecule
through a cell membrane gate having a chemical potential difference
between their two sides. The driving is in this case induced by the
typical open/close mechanism which follow the purely dichotomous
switching largely used in literature \cite{shulten,Millonas,kargol}.

The purpose of our work is to model phenomenologically the possible
physical systems described above. We want to stress here the
qualitative specific results connected to the purely dichotomous
driving.

Thus, differently from the sinusoidal case, no special behavior is
observed in the mean translocation time of the polymer for the case
here studied. However, for this problem, another observable parameter
can be studied. In fact, single molecule experiments are able to
detect and use the instantaneous velocity in order to quantify the
translocation process in forced systems \cite{Bust}.  Remarkably, we
find a non trivial behavior of the polymer translocation velocity as a
function of the mean frequency $\nu$ of the driving with the presence
of a maximum, even if the translocation time shows only a monotonic
behavior. This difference reveals the importance of dealing with
several measures to explore the complex behavior of the polymer
translocation.

The dependence of the stall force $F_{stall}$ of the machine is also
calculated. We find, again, a strong nonmonotonic behavior of
$F_{stall}$ with the frequency, similar to the one found in
\cite{fjf-sine}.

The paper is organized as follows: first we present the model for
polymer and the properties of the stochastic driving force. The main
properties of the translocation process are then calculated:
translocation time, mean velocity and stall force. Finally, we analyze
the dependence of the above properties with the chain stiffness.

\section{The model}

\begin{figure}[htbp]
\centering
\includegraphics[width=8.2cm]{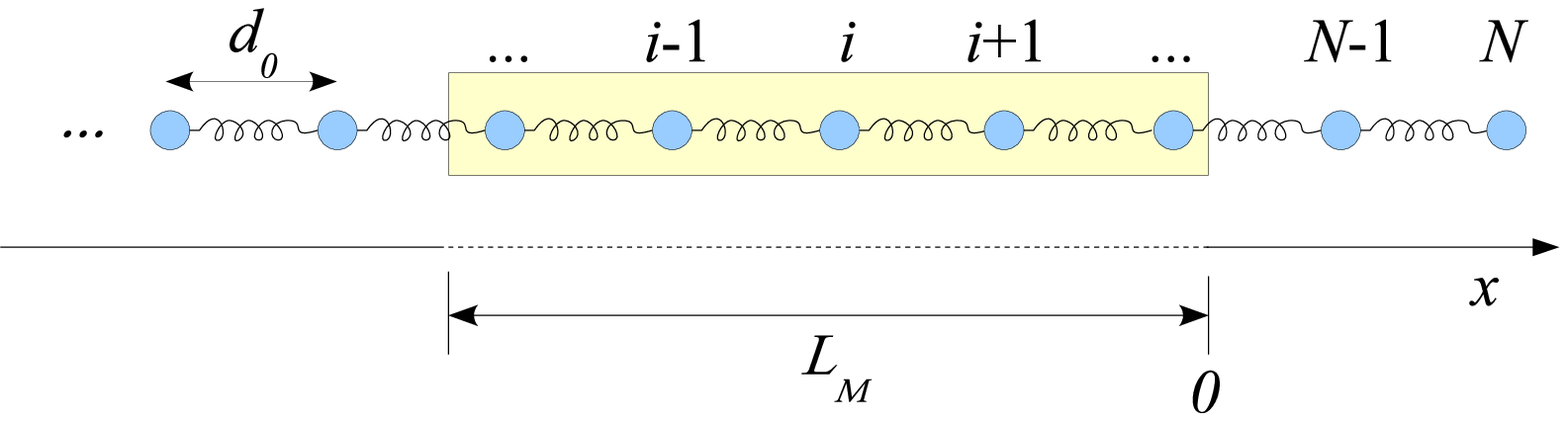}
\includegraphics[angle=-90,width=9.5cm]{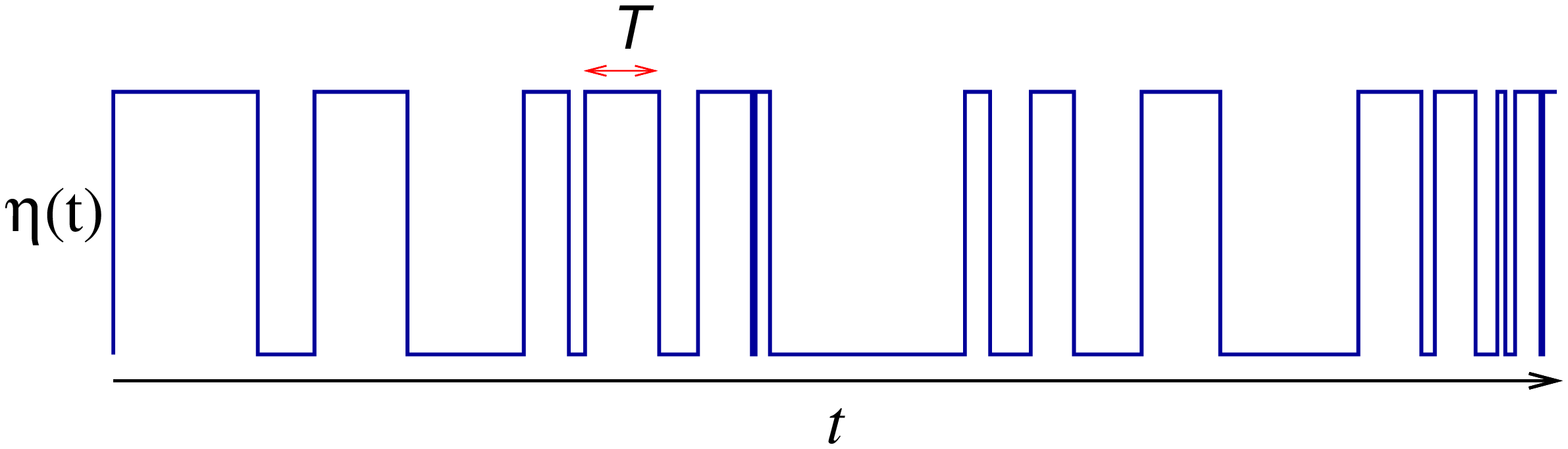}
\caption{Scheme of a linear chain driven by a dichotomous force
  restricted to a small space region (width $L_M$). $T$ is the mean
  time during which the force maintains a same value
  $\{0,F_M\}$.}
 \label{schema}
\end{figure}

The polymer is modeled as a unidimensional chain of $N$
dimensionless monomers connected by harmonic springs \cite{Rouse}.

1d models are suitable in order to describe the dynamics of polymers
constrained to move in confined channels \cite{luan2010}. Also, in
many experimental situations [7] the polymer is stretched, thus
removing the dimensionality dependence of the measured
quantities. Moreover, in this work, we want to fix our attention to
the motor activity in the translocation more than the delay given by
other effects, such as entropic contributions.

The total potential energy is
 \be
 V_{\rm har}=\frac{k}{2}\sum_{i=1}^{N} (x_{i+1}-x_i-d_0)^2,
 \label{v-har}
 \ee
\noindent where $k$ is the elastic constant, $x_i$ the position of the
$i$-th particle, and $d_0$ the equilibrium distance between adjacent
monomers.

The translocation is helped by the presence of a motor which is
activated dichotomously. The machine has a spatial working width $L_M$
and the position $x=0$ represents the right edge of its action (see
Fig.~\ref{schema}). Thus the monomers $i$ such that $x_i \in [-L_M,0]$
experience a force made by the motor. We define $\eta(x,t)$ to
represent the dichotomous force, which fluctuates between two values
$0$ (no force) and $F_M$. Thus \be {\langle \eta(t)
  \rangle}=\frac{F_M}{2} \; \; \, \; {\rm and} \; \; \; {\langle
  \eta(t) \eta(t') \rangle}=\frac{F_M^2}{4}(1+ e^{-2\frac{(t'-t)}{T}})
   \label{eta-corr}
 \ee
Here $T$ gives the mean residence time in each state. With
 respect to the spatial dependence
\be \eta(x) = \left\{   \begin{array} {lr}
     F_M     &    x \in [-L_M,0] \vspace{0.2cm}\\
     0       &    otherwise
   \end{array} \right.
   \label{Potential-s}
 \ee

The dynamics of the $i^{\rm th}$ monomer of the chain is then
described by the following overdamped Langevin equations:
 \be
  \dot{x}_i = -\frac{\partial{V_{har}}}{\partial{x_i}} + \eta(t,x_i) + \xi_{i}(t)
  \label{lang}
 \ee where the viscosity parameter for each monomer is included in the
 normalized time units. $\xi_{i}(t)$ stands for Gaussian uncorrelated
 thermal fluctuation and follows the usual statistical properties
 $\langle\xi_i(t)\rangle=0$ and $\langle\xi_i(t)\xi_j(t')\rangle = 2 D
 \delta_{i j}\delta(t'-t)$.

\section{Results}

We performed a set of $N_{exp}=20,000$ numerical experiment with a
stochastic Runge-Kutta algorithm, using a time step of $dt=0.01$.  The
polymer is compound by $N$ monomers and starts with all the spring at
the rest length ($d_0=1$), and the last monomer of the chain lies at
($x_N=0$), just in the final action range of the dichotomous
force. The noise intensity is held fixed at the value $D=0.001$,
$L_M=5.5$, and $N=12$. The choice of the number of monomers $N$, or
equivalently the length $L$, is arbitrary and this small number has
been used for computational convenience. We note that in a previous
work \cite{fjf-sine}, also with 1d chain, it was found that $\tau$
scales with $L^2=(N-1)^2$. Similarly we find that $v$ scales with
$1/N$.

In this first part, the elastic constant $k$ is held equal to 1, a
meaningful choice that corresponds to a not too rigid approximation
for the polymer. We will study the main observables of the system as a
function of the mean frequency transition $\nu=1/T$.

\subsection{Translocation times}

\begin{figure}[htbp]
\centering
\includegraphics[angle=-90, width=9.5cm]{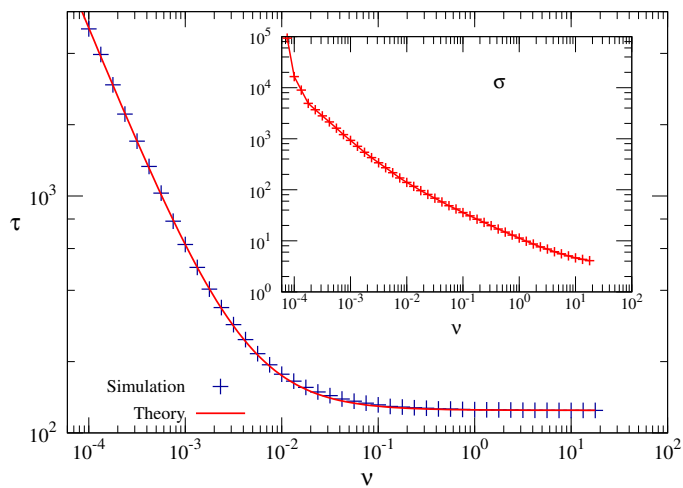}
\caption{Polymer translocation time $\tau$ as a function of the
  frequency $\nu$ of the fluctuating force. The solid line shows the
  theoretical prediction of Eq.~\ref{theor}. In the inset is shown the
  standard deviation of the exit times distribution.}
\label{RA}
\end{figure}

The translocation time $\tau$ is computed as a mean first passage time
of the center of mass of the polymer: the average over the $N_{exp}$
realizations of the time spent by the center of mass of the chain to
reach the position $x=0$.  In Fig.~\ref{RA} we see the value of $\tau$
plotted as function of the mean frequency of the driving, and, in the
inset, the standard deviation $\sigma$ whose values are of the same
order of magnitude than the mean time values, as expected. We find
that $\tau$ is a monotonic function of $\nu$. This result is different
from that for a periodic force (sinusoidal or square wave ones) where
it is observed a minimum in the translocation for $\nu \sim 10^{-2}$
and an oscillating behavior for higher frequencies~\cite{fjf-sine}.

\begin{figure}[htbp]
\centering
\includegraphics[angle=-90, width=9.5cm]{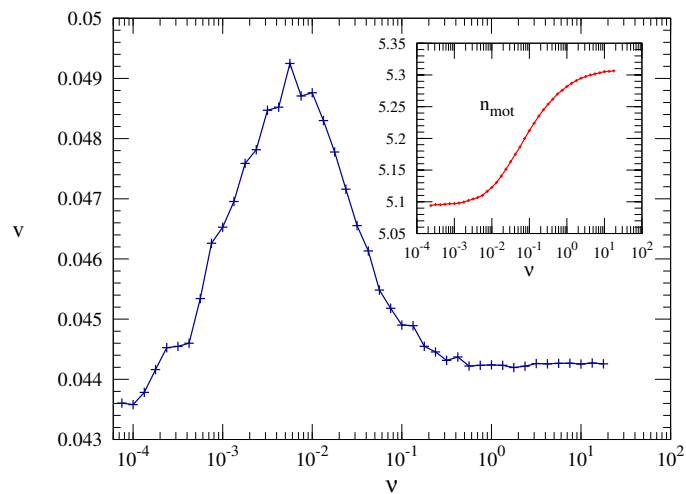}
 \caption{Mean velocity and number of monomers inside the motor while
   in its active state (inset) as a function of the mean frequency
   $\nu$ of the fluctuating force.}
 \label{vRTN}
\end{figure}

In contrast with the behavior of the translocation time, as we will
see, the velocity is not a monotonic function of $\nu$ and a maximum
is found in this function for $\nu \sim 10^{-2}$ (see
Fig.~\ref{vRTN}). Both effects (minimum translocation
time~\cite{fjf-sine} or maximum mean velocity) reveal some interesting similarities with the resonant activation phenomenon~\cite{Pizz2,RA}.

We can make a simple analytical prediction for $\tau(\nu)$ in the low
frequency region which however is found to be valid in a broad
frequency range (see solid line in Fig.~\ref{RA}). Let $\tau_{on}$ be
the value of the exit time when a constant force $F_M$ is applied
during all the dynamics. In the $\nu \to 0$ limit we have to
distinguish between two cases depending on the initial value of the
force, $F_M$ or $0$. In the first case the translocation time is
$\tau_{on}$ corrected in a first approximation by a long waiting time
$T$ if the system switches to the {\em off} state before $\tau_{on}$,
which occurs with a probability $p_s=1-e^{\tau_{on}/T}$. This
correction gives a contribution of $\tau_{on} (1-p_s)+ (T+\tau_{on})
p_s$ to the total time. In the second case there is an additional time $T$ in the off state for escaping. Thus the total translocation time
is
 \be
  \tau \simeq \frac{1}{2}(\tau_{on} + T (1-e^{\tau_{on}/T})) +
 \frac{1}{2} (\tau_{on} + T + T (1-e^{\tau_{on}/T}))
 \label{theor1}
 \ee
Since this equation is derived in the low frequency limit where
$1-e^{\tau_{on}/T} \simeq \tau_{on}/T$ we have
 \be
  \tau \simeq 2 \tau_{on} + T/2
  \label{theor}
 \ee
The intermediate frequency region is characterized by the presence of
the constant force alternated by the absence of the force (diffusive
dynamics) with an average time ratio between them different for
different values of the mean frequency. Surprisingly,
Eq.~(\ref{theor}) also describes in a good way that frequency region.
The third region is instead characterized by a high frequency
switching rate between the two force states. There the translocation
time is much smaller than $T$, the polymer experiences a mean force
$F_M/2$, and $\tau \simeq 2 \tau_{on}$. A careful observation of our
numerical results show that in this high frequency regime
 \be
  \tau \simeq 2 \tau_{on} + T ,
  \label{theorH}
 \ee not observable in Fig.\ref{RA}.
\begin{figure}[htbp]
\centering
\includegraphics[angle=-90, width=9.5cm]{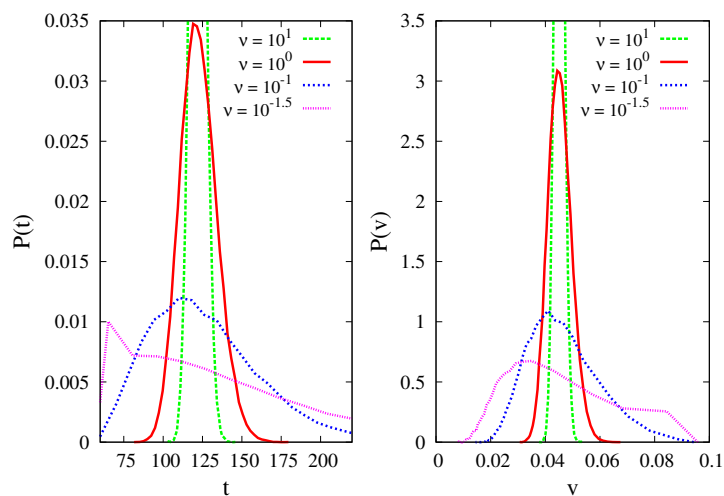}
\caption{First passage time (left) and velocity (right) probability
  distributions at four frequency values.}
\label{P}
\end{figure}

\subsection{Mean velocity}

Fig.~\ref{vRTN} shows the mean velocity $v$ of the polymer as a
function of $\nu$. The inset of the figure shows the average number of
monomers inside the motor during the active states, $n_{mot}$. We can
see that this number is not constant for different values of the mean
frequency $\nu$, at least for the value of the elastic constant $k=1$
used in those calculations.

The main result in the velocity curves is the presence of a well
pronounced maximum, which put in evidence the qualitative difference
between the calculation of the mean first passage time and the mean
velocity. In fact, the velocity is computed as $v_{cm}=1/N_{exp}
\sum_i^{N_{exp}} L/t_i$, where $t_i$ is the escape time in the $i$-th
realization.

As visible in Fig.~\ref{P}, the exit times distribution changes in
shape by changing the mean driving frequency $\nu$. For high values of
$\nu$, the time distribution is very narrow around its mean value
$\tau$. The corresponding probability distribution function for the
velocity is also a narrow function. Decreasing the value of $\nu$, the
distributions are more asymmetric and the width increases. The maximum
of the time distribution moves toward lower values of time, but the
asymmetry changes and higher and higher values of translocation times
are involved. That's why in that region the mean first passage time
$\tau$ increases, although the time of the $P(t)$'s maximum
decreases. The distribution of the velocity change as well; but
because of the increased width of the time distribution, the mean
value of the velocity in that region does not follow the relation
$v_{med}=1/\tau$ and increases with respect to the high frequency
limit, in opposite direction as the one expected from the time
behavior. The reason of this effect is that in the average, the
smaller times have a higher weight in the inverse $1/t_i$ than bigger
ones. Thus the mean velocity increases up to a maximum. Decreasing
$\nu$ in the low frequency region, the average of the times continues
rising up, because the distribution involves higher and higher
times. The velocity, however, now decreases since the very high times
escapes do not contribute importantly to the mean velocity.

In a first approach the translocation velocity in the high frequency
limit is given by \be v_l=\frac{F_M}{2} \frac{n_{mot}}{N}, \ee a
fraction of monomers given by $n_{mot}/N$ experience a force
$F_M/2$. Then, the corresponding translocation time is $2 \tau_{on}$
(remind that $\tau_{on}$ is the escape time if the motor is always
working). On the contrary, in the low frequency limit one half of the
realizations give a very long escape time (and velocity goes to zero)
and another half give $\tau_{on}$. Thus for low frequencies we obtain
the same value of the velocity that for high frequencies.

However, we can see in Fig.~\ref{vRTN} that the low frequency limit of
the mean velocity does not satisfy the relationship just derived,
being lower than the high frequency value $v_l$. This happens because
the force exerted on the polymer is affected by the number of monomers
inside the motor which, as shown in the inset of the figure, also
depends on $\nu$. We will see below a confirmation of the given
relation by using a strong elastic constant between the monomers,
which guarantees a constant number of monomers inside the motor
nevertheless the dynamical conditions are (see Fig.~\ref{ka}).

From Eq.~(\ref{lang}) it is easy to derive the following equation for
the mean velocity \be v=\frac{1}{N} \frac{1}{N_{exp}}
\sum_{i=1}^{N_{exp}} \langle \eta_i(t) \rangle_T = \frac{F_M}{N}
\frac{1}{N_{exp}} \sum_{i=1}^{N_{exp}} \frac{n_{\rm mot, i}^{\rm
    on}(\nu) t^{on}_i} {t_i},
 \label{v_t}
 \ee
where, for each experiment $i$, $n_{\rm mot, i}^{\rm on} (\nu)$ is the
average number of particles inside the motor when the motor is $on$,
$t^{on}_i$ is the total motor working time, and $t_i$ is the
translocation time of each realization.

At high frequency, $t_i^{on}=\frac{1}{2}t_i$. However, decreasing the
frequency for most of the cases, $t_i^{on}>\frac{1}{2}t_i$, during the
translocation the motor spends more time activated that deactivated
since most translocations happen during the activation stage of the
motor. Thus both, the translocation time and the mean velocity
increase\footnote{This is not the case at low values of $k$, where
  the strong change in $n_{mot}$ with the frequency dominates the
  overall behavior and suppress the velocity maximum as shown in
  the inset of Fig.~\ref{ka} for $k=0.1$.}. This behavior changes when $1/\nu \sim
\tau_{on}$. Then $t_i^{on}$ remains constant in Eq.~(\ref{v_t}), $t_i$
increases when $\nu$ decreases and the velocity also decreases towards
the expected $v_l$ value moderate by the mean number of monomers in
the motor in the low frequency limit. This explains the presence of
the maximum in the velocity.

A rough estimation of $n_{\rm mot}$ is given by the fixed value
$n_{\rm mot}=5.5$, corresponding to the distribution of monomer inside
the motor in the case that they maintain the same relative distance,
equal to the rest, over all the dynamics.  This condition will be
completely satisfied for high values of the elastic constant $k$
(rigid chain limit), when $n_{mot}$ becomes independent on $\nu$. As
we will see below, both the high and low frequency limits for the mean
velocity take in that case the same value (see the inset of
Fig.~\ref{ka})

 \[v_{l, theor}= \frac{0.2 \cdot 5.5}{2 \cdot 12}= 0.04583,\]
which is slightly higher than the limit value $v_l$ shown in the inset
of Fig.~\ref{RA} because $n_{mot}(k=1)<5.5$.

\subsection{Stall Force}

The stall force $F_{stall}$ is the force that we need to apply against
the motor in order to stop the polymer translocation. It is a measure
of the strength of the motor and, in this model, it depends on the
frequency of the driving.

\begin{figure}[htbp]
\centering
\includegraphics[width=9.5cm]{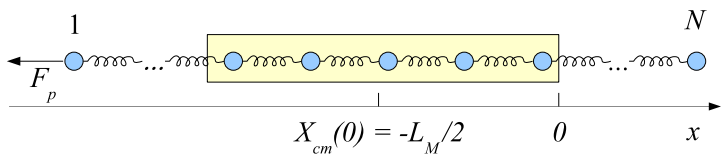}
\caption{A force pull is applied at the first monomer to measure the
  motor stall force.} \label{stall2}
\end{figure}
A set of simulations have been performed by applying a pull force
$F_p$ (see Fig.~\ref{stall2}) on the left extremum of the chain, in
opposite direction to the motor driving force.  The initial condition
for the chain has been fixed with the polymer center of mass in the
center of the motor. Then, the velocity of the center of mass is
measured waiting for the exit on the left or on the right of the motor
region. That way, the force for which the mean velocity is zero gives
$F_{stall}$.

\begin{figure}[htbp]
\centering
\includegraphics[angle=-90, width=9.5cm]{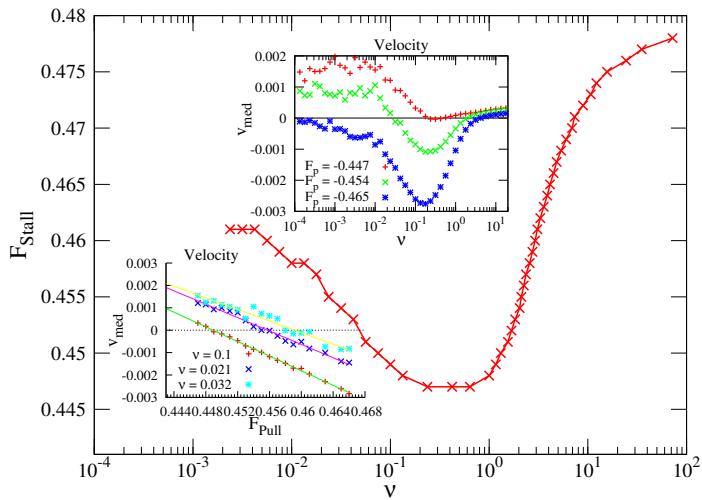}
\caption{Stall force as a function of the frequency $\nu$ of the
  driving.  The upper inset shows the mean velocity as a function of
  $\nu$ for the three pull force values $F_p=0.447, 0.454, 0.465$. The
  bottom  inset shows the linear behavior of the mean velocity as a
  function of the pull force for the three frequency values
  $\nu=0.032, 0.02, 0.1$. The other parameters are the same of
  Fig.~\ref{RA}.}
\label{StallForce}
\end{figure}

Fig.~\ref{StallForce} shows the stall force as a function of the
frequency. As shown in the lower inset, for a given frequency the mean
velocity decreases linearly with $F_p$. The upper inset, shows that
for pull forces of the order of the stall force the velocity presents
a minimum, contrary to the behavior at $F_p=0$ (Fig.~\ref{vRTN}). Then
the stall force, which presents a similar trend, shows a clear minimum
in the same frequency region.

As in the oscillating case \cite{fjf-sine}, the scale variation of the
stall force is small (around $7.5\%$), and an experimental
verification of its behavior with the mean frequency the minimum could
be not immediately simple to perform.

\subsection{Elastic constant dependence}

\begin{figure}[htbp]
\centering
\includegraphics[angle=-90, width=9.5cm]{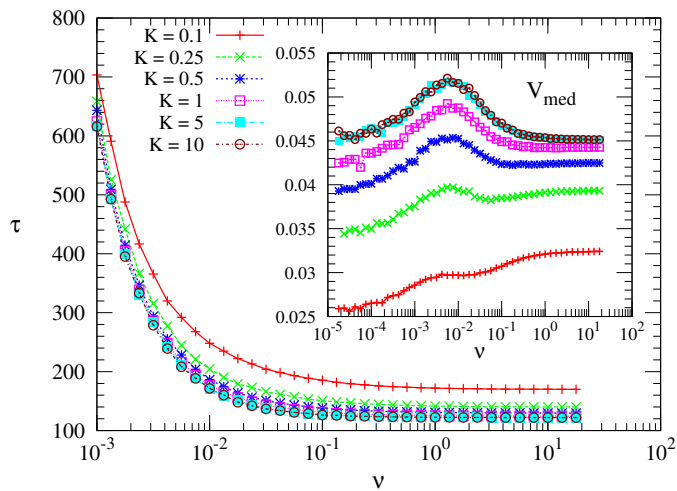}
\caption{Translocation time and mean velocity (in the inset) of the
  polymer chain for different values of the elastic constant $k$.}
\label{ka}
\end{figure}

Finally, we investigate the dependence of translocation time and
velocity on the elastic constant $k$ of the polymer. A magnitude that
strongly depends on $k$ is the mean number of monomer inside the motor
during the pushing cycle, $n_{mot}$.  This number modules the velocity
as it is show in Eq~(\ref{v_t}). Results are plotted in Fig.~\ref{ka},
where the translocation time and velocity (in the inset) are presented
for different value of $k$.
We can see that notable differences (especially visible in the mean
velocity plot) are evident by changing the value of $k$. For the case
$k=0.1$ the velocity looses the maximum, which is always present for
the higher values of $k$. As expected, a clear saturating behavior of
the whole curve is evident by increasing $k$ when the chain behaves
like a rigid bar.  As announced before in the text, in this limit the
mean velocity in the cases of both high and low switching frequency
gives the expected value $v_{l,theor}$ given above. This limit is
already fulfilled for $k=5$.

\section{Conclusions}
The interest in the introduction of simple models is that they can
capture the more relevant features of different processes. In that
way, they can result to be very useful for a coarse-grain description
of different systems.

The model described here studies the translocation process of a
polymer driven by a simple motor which exerts a dichotomous force.  We
analyze the dependence of the translocation time with the mean
frequency of the driving field, and find an analytical expression for
the low frequency regime.  In spite of the monotonic behavior of the
translocation time, the velocity presents a clear maximum at a
resonant value of the mean frequency. We argue that this maximum comes
from the optimization of the "on states" duration of the driving
forces with the corresponding translocation time. The detection of
this maximum, (also seen in the periodic case) could be tackled with
the recent single molecule experimental techniques.

The stall force able to stop the polymer translocation against the
motor has been also evaluated, finding in our calculations results
very close to the oscillating driving, previously studied. The stall
force show a very clear minimum at a resonant mean frequency of the
driving.

The model can have application in artificial nanotechnological devices
driven by dichotomously fluctuating fields, as well as biological pore
membrane with intrinsic noise.

\vspace{0.5cm}
This work has been supported by the project
FIS2008-01240 of the Spain MICINN.

\vspace{0.5cm}

\end{document}